\newcommand{\bee}{\begin{equation}}
\newcommand{\ene}{\end{equation}}
\newcommand{\beea}{\begin{eqnarray}}
\newcommand{\enea}{\end{eqnarray}}

\documentclass[aps,preprint,pre,showpacs,superscriptaddress]{revtex4}
\usepackage{amsmath} %%%% To be able to use 'align' and 'subfigures', etc.
\usepackage{graphicx}% Include figure files
\usepackage{dcolumn}% Align table columns on decimal point
\usepackage{bm}% bold math
\begin{document}
\title{Effect of Dynamic Ions on Band Structure of Plasmon Excitations}
\author{M. Akbari-Moghanjoughi}
\affiliation{Faculty of Sciences, Department of Physics, Azarbaijan Shahid Madani University, 51745-406 Tabriz, Iran}

\begin{abstract}
In this paper we develop a new method to study the plasmon energy band structure in multispecies plasmas. Using this method, we investigate plasmon dispersion band structure of different plasma systems with arbitrary degenerate electron fluid. The linearized Schr\"{o}dinger-Poisson model is used to derive appropriate coupled pseudoforce system from which the energy dispersion structure is calculated. It is shown that the introduction of ion mobility, beyond the jellium (static ion) model with a wide plasmon energy band gap, can fundamentally modify the plasmon dispersion character leading to a new form of low-level energy band, due to the electron-ion band structure mixing. The effects ionic of charge state and chemical potential of the electron fluid on the plasmonic band structure indicate many new features and reveal the fundamental role played by ions in the phonon assisted plasmon excitations in the electron-ion plasma system. Moreover, our study reveals that ion charge screening has a significant impact on the plasmon excitations in ion containing plasmas. The energy band structure of pair plasmas confirm the unique role of ions on the plasmon excitations in many all plasma environments. Current research helps to better understand the underlying mechanisms of collective excitations in charged environment and the important role of heavy species on the elementary plasmon quasiparticles. The method developed in this research may also be extended for complex multispecies and magnetized quantum plasmas as well as to investigation the surface plasmon-polariton interactions in nanometallic structures.
\end{abstract}

\date{\today}

\maketitle
\section{Introduction}

Theories of electron gas excitations are of primary importance in physical properties of solids \cite{kit}. Because of inertialess feature of electrons almost all the electronic properties of matter is influenced by dielectric response of the electron gas to external perturbations \cite{ichimaru1,ichimaru2,ichimaru3}. These properties range from optical and mechanical to transport phenomena \cite{wooten,sarma,stern2,gupta}. In crystalline solids due to complex nature of energy bands the electromagnetic response of the electrons are fundamentally dependent on the energy dispersion and the structure of energy bands in a specific direction \cite{hwang,datta}. One of the well-known features caused by free electrons in metals is the existence of sharp plasmon edge leading to their distinguished optical properties. Plasmons are elementary collective excitations of electrons with many applications in newly emerged interdisciplinary fields such as opto- and nanoelectronics \cite{haug,gardner}, plasmonics \cite{man1,maier} and miniaturized integrated semiconductor circuit industry \cite{markovich} etc. due to their ultrafast terahertz scale electromagnetic response feature \cite{umm,hu1,seeg}. Plasmons also find applications in optical surface engineering due to production of appealing visual effects on nanocoated materials \cite{mlad}. On the other hand, electrons have the first role in many quantum effects in metals and semiconductors such as the Landau quantization leading to the integer quantum Hall and de Haas–van Alphen effects \cite{ash}. Moreover, the role played by electrons in linear and nonlinear wave phenomenon of complex plasmas is tremendous \cite{chen,krall,drazin,stenf1,stenf2,stenf3,stenf4}. Study of electronic properties of solid within the free electron model has a limited applications due to the fact that it ignores interactions between the crystal lattice and the electron gas. Due to strong ionic coupling in solids, in the jellium model, ions usually make the neutralizing positive background and do not contribute to the electronic properties. However, the electrostatic potential of the ionic lattice is the most important source of interaction between electrons and the lattice potential. Moreover, manybody theories such as the density functional theory (DFT) \cite{ethan,white,filinov,bonitz} and Hartree-Fock perturbation \cite{lev} take further step in accounting for interactions among electrons in expense of considerable computational efforts \cite{fro,birdsal,allen}.

In plasma theories, on the other hand, different plasma species are coupled through the common electromagnetic fields that is caused by these species. Therefore, the many body interaction effects is the main building block of the plasma theories. However, in specific oscillation regimes the dynamics effects due to one or more species can be ignored where these species are considered as the jellium in the plasma model. For instance, in a plasma model of dust acoustic excitations both electron and ion dynamics are ignored due to their inertialess character as compared to dust fluid. Despite the long history of pioneering developments \cite{madelung,fermi,hoyle,chandra,bohm,bohm1,bohm2,pines,levine,klimontovich}, recent advancements in quantum kinetic and (magneto)hydrodynamic theories \cite{man2,haasbook,manfredi} have opened new opportunities to explore the fields of complex dense plasmas with arbitrary degree of electron degeneracy and ion coupling strength, ranging from strongly coupled nanometallic compounds and warm dense matter (WDM) \cite{glenzer2,koenig} up to the astrophysical dense objects such as planetary cores and white dwarf stars \cite{chandrasekhar1}. Application of quantum plasma theories have revealed a large number of interesting new linear and nonlinear aspects of collective interactions among different plasma species \cite{se,sten,ses,haas1,brod1,mark1,man3,mold2,sm,fhaas,scripta,kim,mannew,hurst} and has led do discovery of many new phenomena such as resonant shift and electron spill-out \cite{giu}, novel quantum screening \cite{seprl,akbhd2,elak,akbg} and plasmon-soliton \cite{fer}, to name a few. Application of linearized Schr\"{o}dinger-Poisson system for eigenvalue problem has shown many interesting new aspects of quantum plasmas due to the unique dual lengthscale character of plasmon excitations \cite{akbquant,akbheat,akbint,akbnew,akbdirac,akbgap,akbtrans,akbfano}. In current research we would like to extend our previous results of pseudoforce approach to the plasmon energy band structure in complex plasma environments. This may lead to a better understanding of collective effects on electron transport and plasmonic properties of the fourth state of matter which constitutes at least $90$ percent of the visible universe.

\section{The Theoretical Model}

In order to investigate the energy dispersion relation for one-dimensional plasmon excitations we use the linearized Schr\"{o}dinger-Poisson system which is cast into the following normalized coupled pseudoforce model representing the eigenvalue problem for plasmon energies in an ensemble of arbitrary degenerate electron gas in an ambient jellium-like positive background \cite{akbnew}
\begin{subequations}\label{pf}
\begin{align}
&\frac{{d^2{\Psi(x)}}}{{d{x^2}}} + \Phi(x) + E \Psi(x)=0,\\
&\frac{{d^2{\Phi(x)}}}{{d{x^2}}} - {\Psi}(x) = 0,
\end{align}
\end{subequations}
in which $E=(\epsilon-\mu_0)/E_p$ with $E_p=\hbar\sqrt{4\pi e^2 n_0/m_e}$ where $n_0$ and $\mu_0$ are, respectively, the equilibrium number density and chemical potential of the electron gas. Moreover, $m_e$ is the electron mass and $\epsilon$ is the kinetic energy of electrons in the band structure. Also, the normalized functions $\Psi(x)=\psi/\sqrt{n_0}$ and $\Phi(x)=e\phi/E_p$ represent the local probability density and electrostatic energy of the gas, so that, $n(x)=\psi(x)\psi^*(x)$ is the local number density functional. Note that in the Thomas-Fermi approximation the chemical potential remains constant throughout the gas for linear perturbations. At thermal equilibrium for isothermal processes at temperature $T$ the electron number density and chemical potential are connected via a simple equation of state (EoS) \cite{elak}
\begin{equation}\label{iso}
{n} =  - 2{\left( {\frac{m_e}{{2\pi \beta {\hbar ^2}}}} \right)^{3/2}}{\rm{L}}{{\rm{i}}_{3/2}}\left[ { - {\rm{exp}}\left( {\beta \mu_0 } \right)} \right],\hspace{3mm}{P} =  - \frac{N}{\beta }{\rm{L}}{{\rm{i}}_{5/2}}\left[ { - {\rm{exp}}\left( {\beta {\mu_0}} \right)} \right],
\end{equation}
in which $P$ is the statistical pressure satisfying the thermodynamic identity $n=dP/d\mu_0$ and $\beta=1/k_B T$. The polylogarithm function $\rm{Li_\nu}$ is given in terms of the Fermi functions
\begin{equation}\label{li}
{\rm{Li}}_{\nu}( - {{\exp[z]}}) = -\frac{1}{\Gamma (\nu)}\mathop \smallint \limits_0^\infty  \frac{{{x^{\nu-1}}}}{{\exp (x - z) + 1}}{\rm{d}}x,\hspace{3mm}\nu > 0,
\end{equation}
where $\Gamma$ is the ordinary gamma function. The system (\ref{pf}) has simple general solution discussed elsewhere \cite{akbquant}. The linear plasmon dispersion relation is obtained assuming $\Psi(x)=\Psi_1\exp(ikx)$ and $\Phi(x)=\Phi_1\exp(ikx)$ which together with Eq. (\ref{pf}) leads to
\begin{equation}\label{dis1}
\left( {\begin{array}{*{20}{c}}
{E - {k^2}}&1\\
{ - 1}&{ - {k^2}}
\end{array}} \right)\left( {\begin{array}{*{20}{c}}
{{\Psi _1}}\\
{{\Phi _1}}
\end{array}} \right) = \left( {\begin{array}{*{20}{c}}
0\\
0
\end{array}} \right).
\end{equation}
which leads to the simple plasmon dispersion $E=(1+k^4)/k^2$ where $E$ and $k$ are normalized to $E_p$ and $k_p=\sqrt{2m_eE_p}/\hbar$, respectively. The system (\ref{pf}) can be generalized for particle of arbitrary mass $M$ and charge $Z$
\begin{subequations}\label{pfg}
\begin{align}
&\gamma\frac{{d^2{\Psi(x)}}}{{d{x^2}}} - Z \Phi(x) + E \Psi(x)=0,\\
&\frac{{d^2{\Phi(x)}}}{{d{x^2}}} + Z {\Psi}(x) = 0,
\end{align}
\end{subequations}
where $\gamma = m_e/M$ is the fractional mass. The system (\ref{pfg}) the energy dispersion relation $E=(Z^2+\gamma k^4)/k^2$ which in the limit $M=m_e$ and $Z=\pm 1$ reduces to the previous dispersion. It is remarked that the plasmon energy dispersion of electron ($Z=-1$) and positron ($Z=+1$) gases are identical. The plasmon energy dispersion has two distinct lengthscales. For $k\gg 1$ it has particle-like character and reduces to conventional parabolic form $E\simeq \gamma k^2$ while for $k\ll 1$ it has wave-like character and it turns into $E\simeq Z^2/k^2$. The duallength scale character of plasmon excitations, which is evidently due to both single-particle and collective behavior, has shown to produce many fundamental properties in an unmagnetized electron gas \cite{akbheat,akbint,akbnew,akbdirac,akbgap,akbtrans,akbfano}. The dispersion curve has a minimum value at $k_m=\gamma^{1/4}\sqrt{Z}$ and $E_m=2Z\sqrt{\gamma}$ and for $E>E_m$ has two characteristic wavenumbers
\begin{equation}\label{wn}
{k_ \pm } = \sqrt {\frac{{E \pm \sqrt {{E^2} - 4\gamma {Z^2}} }}{{2\gamma }}},
\end{equation}
in which $k_+$ and $k_-$ are particle- and wave-like branches satisfying the complementarity-like relation $k_+k_-=|Z|$. For electron or positron gas we simply have $E_m=k_m=1$ and $k_+=1/k_-$.

\begin{figure}[ptb]\label{Figure1}
\includegraphics[scale=0.67]{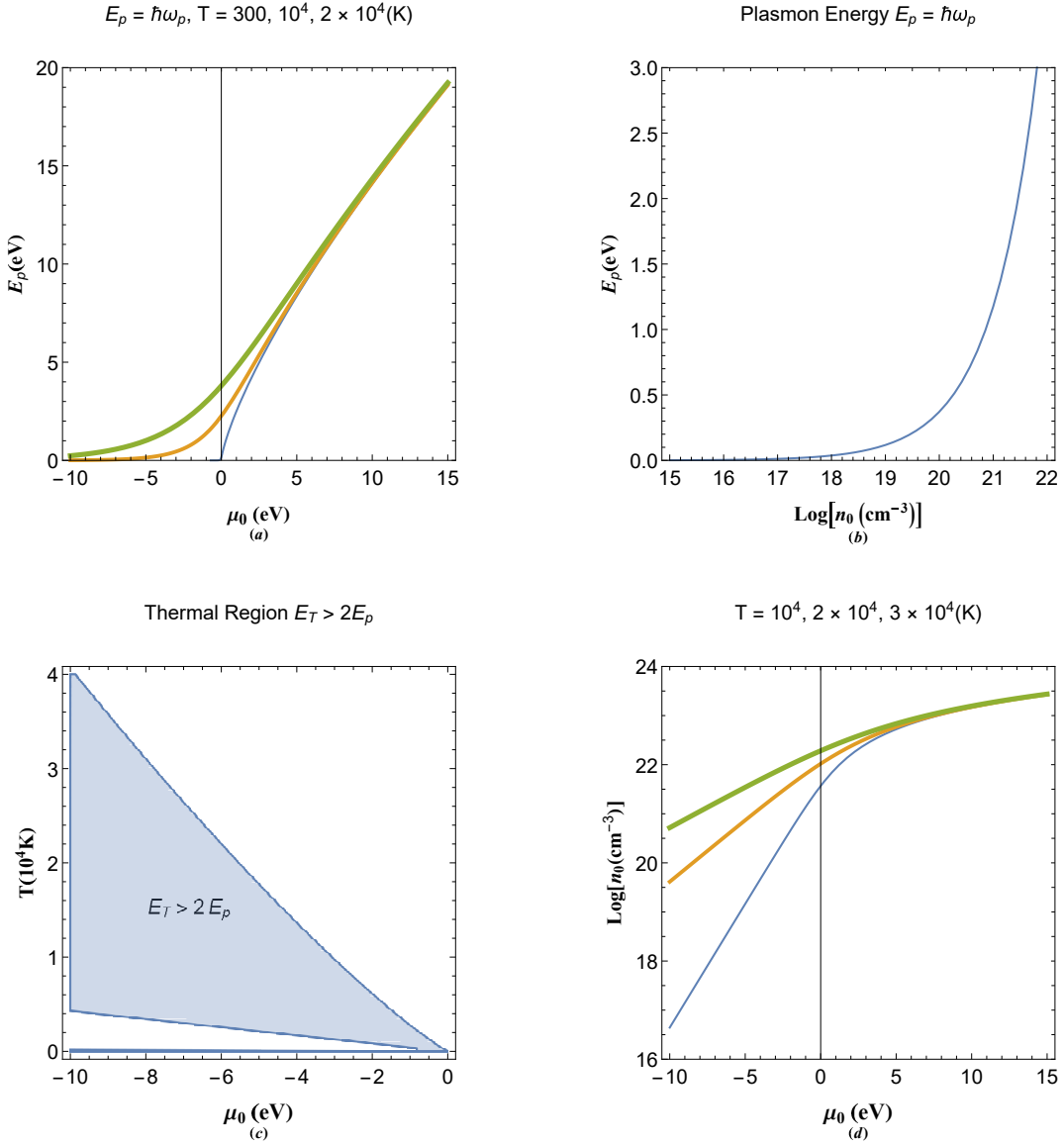}\caption{(a) Variations of electron plasmon energy in terms of the chemical potential of arbitrary degenerate electron gas for different values of electron gas temperature. (b) Variations in plasmon energy in terms of equilibrium electron gas number density. (c) Parametric region in which the thermal energy of electrons exceed twice the plasmon energy of the electron gas. (d) Variations of electron number density versus the chemical potential of the electron gas. The thickness of curves in plots are intended to charcterize the increasing of the values in varied parameter above each panel.}
\end{figure}

In Fig. 1(a) we have shown the values of plasmon energies for different values of the chemical potential of an arbitrary degenerate electron gas for different value of electron gas temperature. In the completely degenerate gas, $\mu_0\gg 1$, or in dilute classical limit, $\mu_0\ll -1$, the effect of electron gas temperature on plasmon energy is insignificant. It is remarked that with increase in the chemical potential for fixed temperature the plasmon energy increases sharply. It is also seen that at room temperature the values of plasmon energy vanishes for a classical electron gas where the chemical potential becomes negative. This result is better depicted in Fig. 1(b) where the variations of plasmon energy is compared in terms of the number density in logarithmic scale. The plasmon energy increases sharply as the electron number density increases beyond the degeneracy limit $n_0\simeq 10^{18}$cm$^{-3}$. Furthermore, Fig. 1(c) the region in temperature-chemical potential for which the thermal energy in the gas is comparable to twice as much as the plasmon energy (which is the plasmon gap studied in the following). It is remarked that for dilute classical electron gas $\mu_0<0$ at low temperature there is small ribbon-shaped area satisfying the condition. However, for degenerate electron gas such as for semiconductors and metals the thermal energies never reach the value $2E_p$ even for very large electron temperatures. The later is because the increase in the temperature of the gas also increases its plasmon energy. However, in the classical region $\mu_0<0$ there is region of high temperature area in which $E_T\ge 2E_p$. Figure 1(d) shows the variation of the chemical potential versus the electron number density for different values of temperature. It ia again remarked that for fully degenerate gas the effect of temperature becomes insignificant. However, for the classical electron gas for a fixed chemical potential increase of the temperature significantly increases the electron number density.

\begin{figure}[ptb]\label{Figure2}
\includegraphics[scale=0.67]{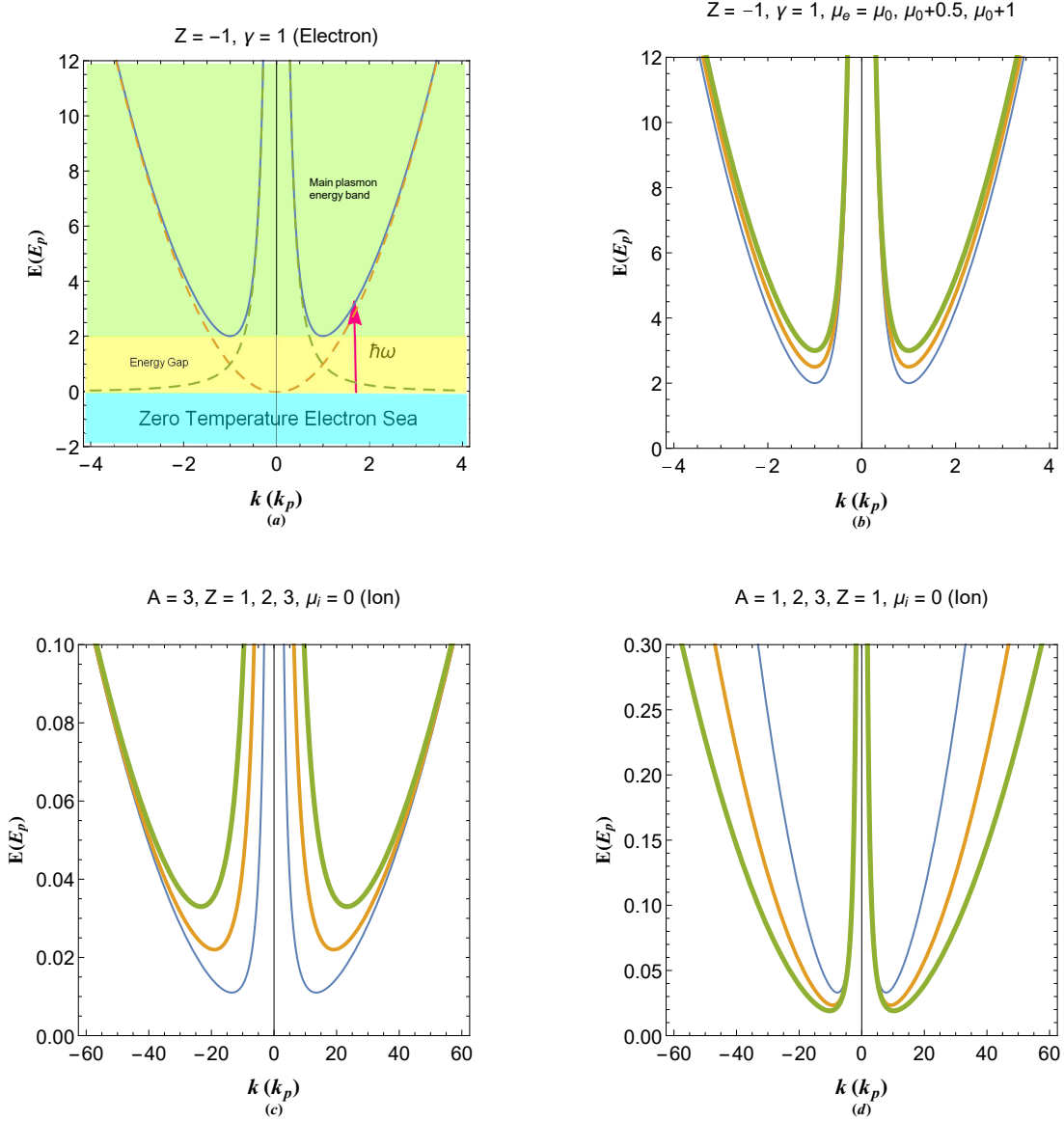}\caption{(a) The plasmon energy band struture of plasmon excitations electron gas of arbitrary degeneracy in Thomas-Fermi approximation in jellium model. (b) Variations in the electron plasmon band structure with changes in the normalized chemical potential. (c) Plasmon dispersion structure for heavy particle and the effects of particle charge state on the band structure. (d) Effect of atomic number of ion species on the energy band structure of neutralized ion fluid. The thickness of curves in plots are intended to charcterize the increasing of the values in varied parameter above each panel.}
\end{figure}

Figure 2 shows the variations in plasmon energy dispersion curves for various parameters. Figure 2(a) is the dispersion curve for electron gas with arbitrary degree of degeneracy. The energy level $E=0$ corresponds to top of the electron see at zero temperature ($\epsilon=\mu$). At the full degeneracy limit ($\mu=\epsilon_F$ with $\epsilon_F$ being the Fermi energy of the gas) all available electron energies are packed in the area $E\le 0$ or $\epsilon<\epsilon_F$ ($\epsilon_F\simeq 1.59$eV for cesium). The solid curves indicate the main plasmon band and all excited plasmon energies reside for $E>2$. The dashed curve, on the other hand, shows the free electron dispersion. The origin of plasmon band gap is in fact a resonant quantum scattering between free electron-like ($E=k^2$)  propagations by their collective excitation with the characteristic dispersion of $E=1/k^2$, shown as dashed curves in Fig. 1(a). At low temperature where thermal energy of electrons is low (room temperature thermal energy $\simeq 0.026$meV) only a small fraction of electron in top of electron Fermi sea may attain kinetic energies $\epsilon>2E_p$ ($E_p\simeq 2.9$eV for cesium) in order to reach the plasmon conduction band. The photon frequencies to excite such electrons must be at least twice that of the electron plasmon frequency. This small rate of excitation take place in the form of electron-hole excitations. However, as the temperature rises more and more electron reach the plasmon band and plasmonic conduction may be possible. One important feature of the band structure is the effective mass of electron at the specific scattering wavenumber $k$. For excitations with $k\gg 1$ (for wavenumbers much larger compared to plasmon wavenumber of the electron gas) the effective mass $m_b=m^*/m_e=2(d^2E/dk^2)^{-1}=k^4/(k^4+3)$ (in the normalized form) which is definitely positive in the whole plasmon band approaches that of the free particle limit, i.e. $m^*\simeq m_e$. On the other hand, for $k\ll 1$ one obtains $m_b\to 0$. Therefore, to distinct regimes of fast wave-like ($\omega>2\omega_p$ and $k<k_p$) and slow particle-like ($\omega>2\omega_p$ and $k>k_p$) electron response to external perturbations are possible in the plasmon band. Note that in room temperature there is a very small probability for plasmon excitations at the Fermi surface of metals due to limiting plasmon occupation number factor $f(E,\theta)=[\exp(2E/\theta)+1]^{-1}$ ath thermal equilibrium, where $\theta=T/T_p$ with $T_p=E_p/k_B$ being the plasmon temperature.

Figure 2(b) shows the effect of chemical potential on the electron plasmon dispersion. It is remarked that as the electron gas become dense at constant temperature $E_m$ increases but $k_m$ does not change. For a classical dilute electron gas the chemical potential can become negative and the plasmon energy can have small values comparable to room temperature thermal energy in which case the bottom of plasmon conduction band sink into the electron sea. The later is the reason for why plasmon oscillations of electron gas are dominant for classical rather than degenerate electron gas (with fixed jellium-like positive background). Figure 2(c) shows plasmon dispersion curves for classical gass of charged particle with atomic number $A$ ($\gamma\simeq 1/2Am_p$ with $m_p$ being the proton mass) and different charge values, $Z$. The values of $E_m\ll 2E_p$ and $k_m\gg k_p$ are due to very small $\gamma$ values of ions compared to the electron. The in the very high wavenumber limit the dispersion approaches the particle-like branch $\gamma k^2$. It is remarked that as the charge value increases $k_m$ decreases but $E_m$ increases significantly. Such changes in the charge alters the wave-like branch but does not significantly modify the particle-like branch of the dispersion. It is concluded that the ions charge state significantly affects the wave-like behavior of plasmon excitations. On the other hand, Fig. 2(d) shows the effect of changes in the atomic number of ionic gas on the dispersion pattern. It is seen that in this case the effect is prounanced on the particle-like branch while the wavelike branch stays almost intact. It is also remarked that with in crease in the value of $A$, $k_m$ increases while $E_m$ decreases as opposed to the variations in Fig. 2(c).

\section{Plasmons in Dynamic-Ion Environments}

The model in Sec. III is known as the jellium model in which neutralizing background ions are static. In this section we will explore how the ion dynamics even negligible compared to that of electrons can fundamentally modify the plasmon dispersion character in electron-ion plasma. To this end, we consider the pseudoforce system
\begin{subequations}\label{pfei}
\begin{align}
&\frac{{d^2{\Psi_1(x)}}}{{d{x^2}}} + \Phi(x) + E \Psi_1(x)=0,\\
&\gamma \frac{{{d^2}{\Psi _2}(x)}}{{d{x^2}}} - Z\Phi (x) + (E + \mu ){\Psi _2}(x) = 0,\\
&\frac{{d^2{\Phi(x)}}}{{d{x^2}}} - {\Psi_1}(x) + Z {\Psi_2}(x) = 0,
\end{align}
\end{subequations}
where $\Psi_1$ and $\Psi_2$ respectively denote the probability density of electrons and ions in the system and we assumed the ions are classical $\mu_i\simeq 0$. Solving the system for plasmon eigenenergy values leads to the following dispersion relation
\begin{equation}\label{disei}
\begin{array}{*{20}{c}}
{E_{\pm} = \frac{1}{{2{k^2}}}\left[ {\lambda  \pm \sqrt {{\lambda ^2} + 4{k^4}\left( {{k^4} - \lambda  - \gamma  + 1} \right) + 4\mu{k^2} } } \right]}\\
{\lambda  = \left( {1 + \gamma } \right){k^4} - \mu {k^2} + {Z^2} + 1},
\end{array}
\end{equation}
It is evident that the dispersion relation has two distinct branches due to dual species gas and reduces to dispersion relation for electron gas in jellium background for $\gamma=Z=0$. The plasmon dispersion of electron-ion plasma gives rise to upper and lower excitation bands.

\begin{figure}[ptb]\label{Figure3}
\includegraphics[scale=0.67]{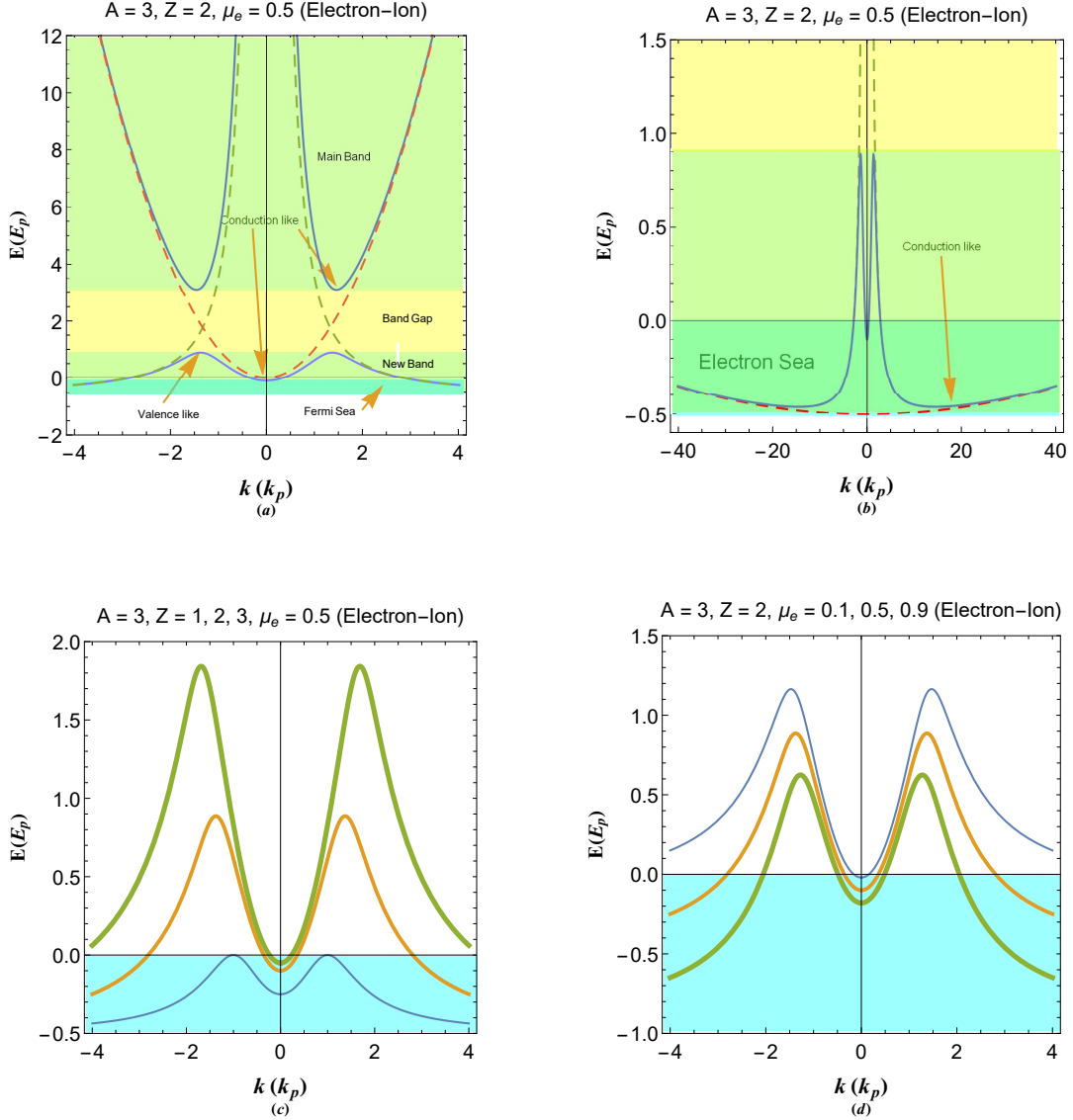}\caption{(a) Plasmon energy band for electron-ion plasma which dynamic ions and arbitrary degenerate electron gas. (b) Phonon assisted plasmon conduction band in a wider scale for given plasma parameters. (c) Effect of dynamic ion charge on the plasmon band structure of electron-ion plasma. (d) Effect of the electron gas chemical potential on the energy dispersion on electron ion plasmas. The thickness of curves in plots are intended to charcterize the increasing of the values in varied parameter above each panel.}
\end{figure}

This is shown in Fig. 3(a) together with the parabolic free electron dispersion $k2$ and ion plasmon dispersion $E=(Z^2+\gamma k^4)/k^2-\mu$ shown as the dashed asymptotic curves. Note that in each plot the point $E=0$ denotes $\epsilon=\mu$ where $\epsilon$ is the normalized kinetic energy of electrons and $E=-\mu$ denotes the zero kinetic energy point. It is remarkable that introduction of mobile ions fundamentally alters the plasmon band structure adding a lower band overlapping the electron sea. It is evident that the new band originates due to the coupling of free electron and ion plasmon excitations coupling (crossing of the dashed dispersion curves). Physically, this is interpreted as the resonant scattering of electrons by the wavelike branch of ion plasmon excitations. The latter effect is completely different from the band gap structure caused by the resonant scattering of free electrons by the lattice potential of static ions in crystalline materials. Moreover, the main band is slightly lifted so that the height of the band gap remains almost unchanged. The great resemblance of current band structure to that in crystal lattices \cite{kit} is remarkable. The new band contains a valence-like ($m_b<0$) structure around $k\simeq 1$ as well as two conduction like ($m_b>0$) valleys at $k<1.5$ and $k>1.5$. The lower energy band can have a significant effect on low temperature variation of macroscopic quantities such as the specific heat and current density. Note that electrons in valence- and conduction-like band response oppositely to the external field due to the relation $a=q{\cal E}/m_e$ where $a$ and ${\cal E}$ refer to the acceleration and external electric field respectively. Note also that valence bands contain $k$ ranges for which the plasmon group-speed $v_g=(1/\hbar)dE/dk$ can be either positive or negative. Due to different curvature values of the conduction-like valleys the electron at lower $k$ values respond faster compared to those in large $k$ plasmon conduction electrons. It is also noted that for very low-wavenumber (very large wavelength) plasmon excitations electrons almost feel free. Because the new band sinks into the electron sea, the electrons need not to be thermally excited in order to contribute collectively. However, there is zero-temperature cut-off wavenumber range $0.5< k < 3$ for plasmon excitations with electron chemical potential value $\mu_0=\mu_0/2E_p=0.5$, $A=3$ and $Z=2$. Figure 3(b) shows the plasmon dispersion band structure of electron-ion plasma at a larger wavenumber scale. It is seen that the new energy band approaches the particle-like branch of ion $E=\gamma k^2-\mu$ shown as dashed curve in Fig. 3(b). Note that in large wavenumber plasmon excitation in conduction band electrons effective mass equal nearly that of the ions. The later phonon-coupled plasmon excitations are less effective in the electronic properties of the plasma compared to the large wavelength plasmon excitations. Figure 3(c) and 3(d) reveal the effects of ion charge state and elctron fluid chemical potential on the new band structure in electron-ion plasma. It is seen that with increase in the ion charge the wavenumber cut-off range decreases and more and more electrons in the Fermi gas contribute to the plasmon excitations at lower temperatures. Figure 3(d) on the other hand reveals that for very low values of $\mu=0.1$ very few electrons contribute to plasmon excitations while increase of $\mu$ (consequently the degree of degeneracy) increases the portion of electrons and wavenumbers in the plasmon excitations. The later is one of the fundamental feature of metal behavior formation in solids which is beautifully illustrated in current plasmon band structure of dense electron-ion plasmas.

\section{Effect of Ion Charge Screening}

Let us now study the effect of charge screening on the plasmon band structure of electron-ion plasmas. In doing this we consider the following generalized pseudoforce system
\begin{subequations}\label{dpf}
\begin{align}
&\frac{{{d^2}\Psi_1 (x)}}{{d{x^2}}} + 2\xi\frac{{d\Psi_1 (x)}}{{dx}} + \Phi (x) + {E}\Psi_1 (x) = 0,\\
&\gamma \frac{{{d^2}{\Psi _2}(x)}}{{d{x^2}}} + 2\gamma\xi\frac{{d\Psi_2 (x)}}{{dx}}- Z\Phi (x) + (E + \mu ){\Psi _2}(x) = 0,\\
&\frac{{{d^2}\Phi (x)}}{{d{x^2}}} + 2\xi\frac{{d\Phi (x)}}{{dx}} - \Psi_1 (x) + Z\Psi_2 (x) = 0,
\end{align}
\end{subequations}
where $\xi=k_{sc}/k_p$ with the normalized screening parameter $\xi^2 = ({E_p}/2{n_0})\partial n/\partial \mu  = (1/2\theta ){\rm{L}}{{\rm{i}}_{1/2}}\left[ { - \exp ({2\mu}/\theta )} \right]/{\rm{L}}{{\rm{i}}_{3/2}}\left[ { - \exp (2{\mu}/\theta )} \right]$ being the one-dimensional screening wavenumber in the finite temperature Thomas-Fermi model \cite{akbnew}. The eigenenergy equation for the system (\ref{dpf}) may be found using the transformations $\Psi_1(x)=\psi_1(x)\exp(-\xi x)$, $\Psi_2(x)=\psi_2(x)\exp(-\xi x)$ and $\Phi_1(x)=\phi_1(x)\exp(-\xi x)$ and using $\psi_1(x)=\psi_{11}\exp(ikx)$, $\psi_2(x)=\psi_{21}\exp(ikx)$ and $\phi(x)=\phi_{1}\exp(ikx)$. Then we find
\begin{equation}\label{dis2}
\left( {\begin{array}{*{20}{c}}
{E - {k^2} - {\xi ^2}}&0&1\\
0&{E + \mu  - \gamma ({k^2} + {\xi ^2})}&{ - Z}\\
{ - 1}&Z&{ - {k^2} - {\xi ^2}}
\end{array}} \right)\left( {\begin{array}{*{20}{c}}
{{\psi _{11}}}\\
{{\psi _{21}}}\\
{{\phi _1}}
\end{array}} \right) = \left( {\begin{array}{*{20}{c}}
0\\
0\\
0
\end{array}} \right).
\end{equation}
The dispersion relation is then given as
\begin{equation}\label{diseid}
\begin{array}{*{20}{c}}
{{E_ \pm } = \frac{1}{{2{K^2}}}\left[ {\Lambda  \pm \sqrt {{\Lambda ^2} + 4{K^4}\left( {{K^4} - \Lambda  - \gamma  + 1} \right) + 4\mu {K^2}} } \right]}\\
\begin{array}{l}
\Lambda  = \left( {1 + \gamma } \right){K^4} - \mu {K^2} + {Z^2} + 1,\\
K = \sqrt {{k^2} + {\xi ^2}}.
\end{array}
\end{array}
\end{equation}
It is obvious to find out that all the dispersion relations we have calculated so far do not depend on the sign of the ion charge. The Eq. (\ref{diseid}) reduce to that of the unscreened jellium model for $\gamma=\xi=Z=0$. Note also that in the limit $\xi=0$ we retain the dispersion for unscreened electron-ion plasma (\ref{disei}). It is evident that there are two distinct bands in the dispersion structure similar to the previous case.

\begin{figure}[ptb]\label{Figure4}
\includegraphics[scale=0.66]{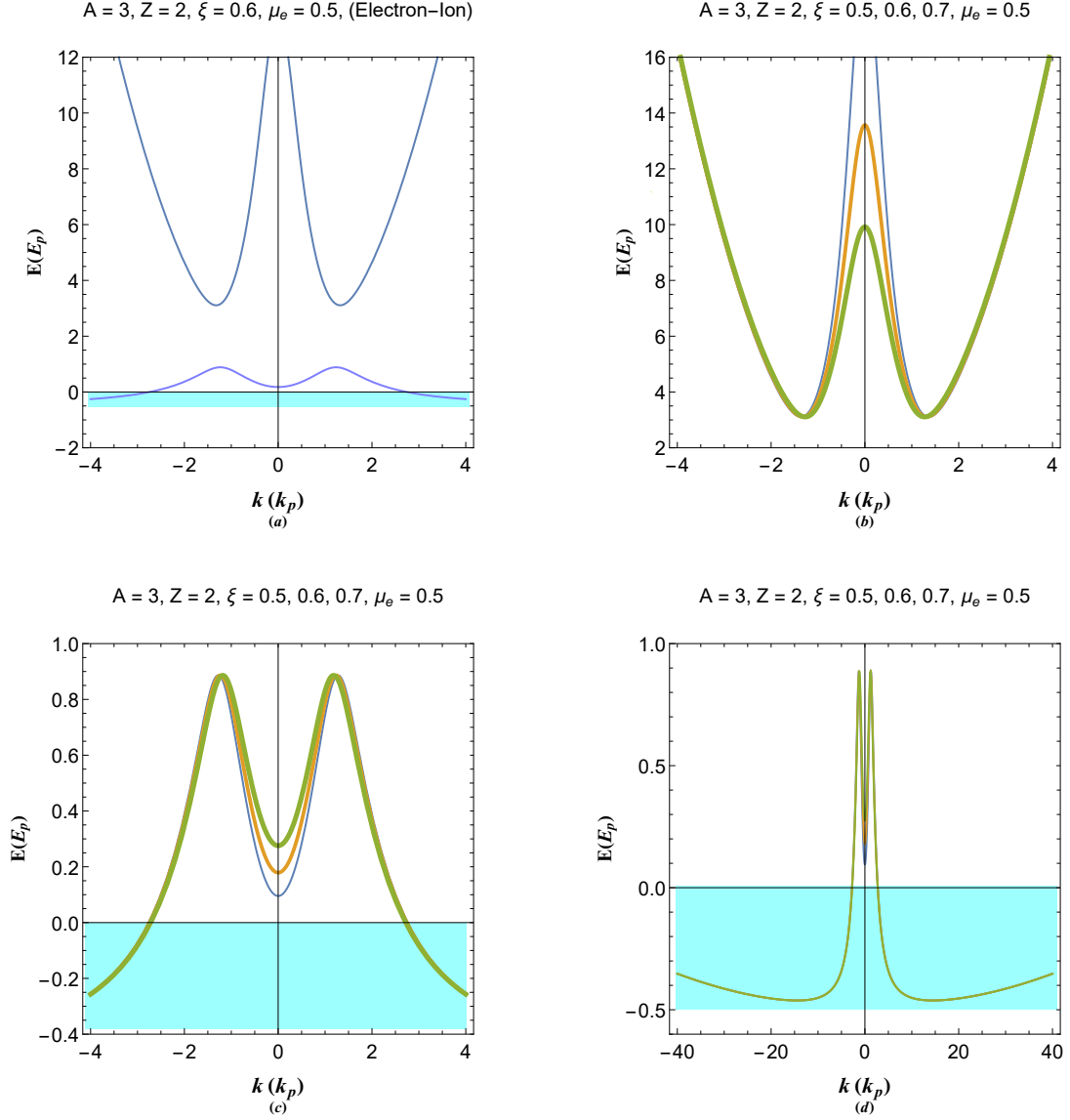}\caption{(a) The plasmon energy band structure of electron-ion plasma with screened ions. (b) The effect of normalized charge screening parameter on the main plasmon conduction band for given plasma parameters. (c) Effect of normalized charge screening parameter on lower plasmon conduction band for high wavelength excitations. (d) Effect of normalized charge screening parameter on lower plasmon conduction band for small wavelength excitations. The thickness of curves in plots are intended to charcterize the increasing of the values in varied parameter above each panel.}
\end{figure}

Figure 4 shows the energy dispersion of plasmon excitations in screened electron-ion plasma. Figure 4(a) as compared to Fig. 3(a) reveals that introduction of charge screening reduces electron population in small wavenumber range of plasmon conduction band. The details of dispersion curve variations for each band are shown in Figs. (c)-(d). It is seen from Fig. 4(b) that the in screened plasma the wave-like branch in main dispersion band is limited. It is also remarked from Fig. 4(c) that increase in the value of screening parameter leads to shift of low wavenumber conduction valley to higher values. It is obvious that such a shift significantly lowers the plasmon states in the lower energy band which contributes the most to the plasmon excitations in electron-ion plasmas. Moreover, Fig. 4(d) shows that the charge screening has insignificant effect on the large wavenumber conduction valley in the lower plasmon band.

\section{Plasmon Dispersion in Pair Plasmas}

For the sake of completeness we would like to investigate the plasmon band structure in pair plasmas. For instance in a electron-pair-ion plasma the pseudoforce system reads
\begin{subequations}\label{pfei}
\begin{align}
&\frac{{d^2{\Psi_1(x)}}}{{d{x^2}}} + \Phi(x) + E \Psi_1(x)=0,\\
&\gamma \frac{{{d^2}{\Psi _2}(x)}}{{d{x^2}}} - Z\Phi (x) + (E + \mu ){\Psi _2}(x) = 0,\\
&\gamma \frac{{{d^2}{\Psi _3}(x)}}{{d{x^2}}} + Z\Phi (x) + (E + \mu ){\Psi _3}(x) = 0,\\
&\frac{{d^2{\Phi(x)}}}{{d{x^2}}} - {\Psi_1}(x) + Z {\Psi_2}(x) - Z {\Psi_3}(x)= 0.
\end{align}
\end{subequations}
The dispersion relation then reads
\begin{equation}\label{diseid2}
\begin{array}{*{20}{c}}
\begin{array}{l}
{E_0} = \gamma {k^2}-\mu \\
{E_ \pm } = \frac{1}{{2{k^2}}}\left[ {\lambda  \pm \sqrt {{\lambda ^2} - 4{k^4}\left( {\gamma{k^4}  + 2{Z^2} + \gamma } \right) + 4{k^2}\left( {1 + {k^4}} \right)} } \right]
\end{array}\\
{\lambda  = \left( {1 + \gamma } \right){k^4} - \mu{k^2}+ 2{Z^2} +1 }.
\end{array}
\end{equation}
It is evident that there are three bands corresponding to three species in the plasma.

\begin{figure}[ptb]\label{Figure5}
\includegraphics[scale=0.67]{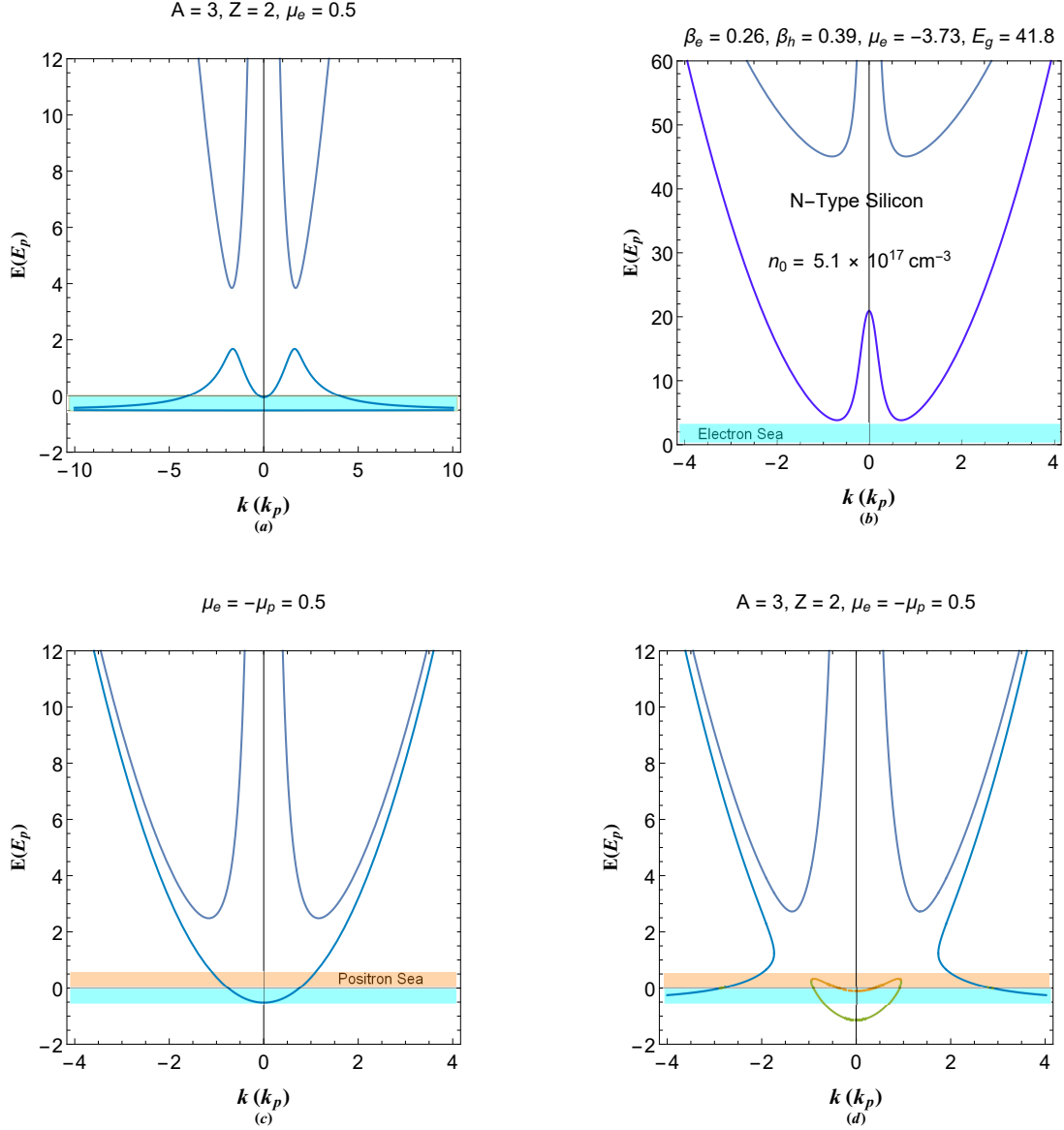}\caption{(a) Plasmon energy dispersion of pair-ion electron-ion plasmas. (b) Plasmon dispersion structure of electron-hole pair with static ions in intrinsic semiconductor plasmas. Plasmon excitation band structure of electron-positron plasma (c) without and (d) with dynamic ion species.}
\end{figure}

The dispersion band structure for (\ref{diseid2}) is shown in Fig. 5(a). It is remarked that the new band $E=\gamma k^2$ which is the parabolic dispersion of ion now appears as the new wide band. Due to very large effective mass of electrons associated with this new conduction band it does not significantly contribute to electronic properties of electron plasmon excitations. Another instance of pair plasma is the electron-hole plasma relevant to semiconductors. The pseudoforce system reads
\begin{subequations}\label{ehi}
\begin{align}
&\gamma_e \frac{{{d^2}{\Psi _e}(x)}}{{d{x^2}}} + \Phi (x) + E{\Psi _e}(x) = 0,\\
&\gamma_h \frac{{{d^2}{\Psi _h}(x)}}{{d{x^2}}} - \Phi (x) + (E - E_g){\Psi _h}(x) = 0,\\
&\frac{{d^2{\Phi(x)}}}{{d{x^2}}} - {\Psi_e}(x) + {\Psi_h}(x)= 0,
\end{align}
\end{subequations}
where $\Psi_e$, $\Psi_h$ and $\Psi_i$ refer to probability density of electron, hole and static ion species. Also, $\gamma_e=m_e/m^*_e$, $\gamma_h=m_e/m^*_h$ and $E_g=\mu_h-\mu_e$ \cite{ae} is the normalized (to $E_p$) gap energy of semiconductor \cite{ae}. The characteristic eigenenergy equation in this case follows
\begin{equation}\label{dis3}
\left( {\begin{array}{*{20}{c}}
{E - {\gamma _e}{k^2}}&0&1\\
0&{E - {E_g} - {\gamma _h}{k^2}}&{ - 1}\\
{ - 1}&1&{ - {k^2}}
\end{array}} \right)\left( {\begin{array}{*{20}{c}}
{{\psi _{e1}}}\\
{{\psi _{h1}}}\\
{{\phi _1}}
\end{array}} \right) = \left( {\begin{array}{*{20}{c}}
0\\
0\\
0
\end{array}} \right).
\end{equation}
Note that the ions have chosen to be static in this case. The corresponding plasmon energy dispersion reads
\begin{equation}\label{ehdis}
\begin{array}{*{20}{c}}
{{E_ \pm } = \frac{1}{{2{k^2}}}\left\{ {\lambda  \pm \sqrt {{{\left[ {{E_g} + \left( {{\gamma_h} - {\gamma_e}} \right){k^2}} \right]}^2}{k^4} + 4} } \right\}},\\
{\lambda  = \left( {{\gamma_e} + {\gamma_h}} \right){k^4} + {E_g}{k^2} + 2},
\end{array}
\end{equation}
where $\beta_e=0.26=1/\gamma_e$ and $\beta_h=0.39=1/\gamma_h$ are the effective-mass ratios for electrons and holes in the silicon semiconductor, respectively, which we used in Fig. 5(b). The dispersion curve is shown in the figure for an electron doped silicon with room temperature number-density of $n_0\simeq 5\times 10^{17}$cm$^{-3}$ ($E_p\simeq 27$meV slightly larger than the electron thermal energy) consisting of two distinct branches. Note that the energy values shown above the plot are normalized to plasmon energy. In fact the unscaled chemical potential value for this density is $\mu_e\simeq -0.1$eV and the unscaled gap energy is $\epsilon_g\simeq 1.12$eV. The valence-like plasmon band is not present in this case. It is seen that thermal excitations are needed to excite electron to plasmon conduction bands. Furthermore, it is found that the upper plasmon band is far higher to be reached by electron sea via thermal excitations. The lower conduction band can be populated mostly with electrons by small thermal agitations so that doped silicon plasma is N-type in this case. The pool of electron holes is in negative side of the axis and extends to $E=E_g+\mu_e$ with $E=0$ denoting the Fermi energy level which is closer to the conduction band for N-type semiconductors, as is the case. For the electron-positron plasma we have
\begin{subequations}\label{ep}
\begin{align}
&\frac{{{d^2}{\Psi _e}(x)}}{{d{x^2}}} + \Phi (x) + E{\Psi _e}(x) = 0,\\
&\frac{{{d^2}{\Psi _p}(x)}}{{d{x^2}}} - \Phi (x) + (E + 2\mu_e ){\Psi _p}(x) = 0,\\
&\frac{{d^2{\Phi(x)}}}{{d{x^2}}} - {\Psi_e}(x) + {\Psi_p}(x)= 0,
\end{align}
\end{subequations}
where we have used $\mu_e+\mu_p\simeq 0$. This leads to the dispersion relation
\begin{equation}\label{epdis}
{E_ \pm } = ({k^4} \pm \sqrt {{\mu_e ^2}{k^4} + 1}  + 1)/{k^2}.
\end{equation}
The plot of dispersion branches shows that in electron-positron pair plasma there is significant overlap between the electron sea and the small wavenumber band available for electron-positron plasmon excitations. This pair plasma is zero band gap plasmonic system with nearly a free electron parabolic dispersion conduction band. In the following it may be illustrative to consider a more realistic electron-positron pair plasma in the presence of a dynamic ion species. The pseudoforce system in this case is
\begin{subequations}\label{ep}
\begin{align}
&\frac{{{d^2}{\Psi _e}(x)}}{{d{x^2}}} + \Phi (x) + E{\Psi _e}(x) = 0,\\
&\frac{{{d^2}{\Psi _p}(x)}}{{d{x^2}}} - \Phi (x) + (E + 2\mu_e ){\Psi _p}(x) = 0,\\
&\gamma\frac{{{d^2}{\Psi_i}(x)}}{{d{x^2}}} - Z \Phi (x) + (E + \mu_e ){\Psi _i}(x) = 0,\\
&\frac{{d^2{\Phi(x)}}}{{d{x^2}}} - {\Psi_e}(x) + {\Psi_p}(x) + Z{\Psi_i}(x)= 0,
\end{align}
\end{subequations}
with a long expression for the dispersion solution which is avoided here for simplicity. The characteristic eigenenergy equation in this case follows
\begin{equation}\label{dis4}
\left( {\begin{array}{*{20}{c}}
{E - {k^2}}&0&0&1\\
0&{E - {k^2} + 2{\mu _e}}&0&{ - 1}\\
0&0&{E - \gamma {k^2} + {\mu _e}}&{ - Z}\\
{ - 1}&1&Z&{ - {k^2}}
\end{array}} \right)\left( {\begin{array}{*{20}{c}}
{{\psi _{e1}}}\\
{{\psi _{p1}}}\\
{{\psi _{i1}}}\\
{{\phi _1}}
\end{array}} \right) = \left( {\begin{array}{*{20}{c}}
0\\
0\\
0\\
0
\end{array}} \right).
\end{equation}
The dispersion curve is shown in Fig. 5(d). The complex structure shows multiple band structure with the superposition of asymptotic previously studied dispersion curves, namely, the electron-positron (Fig. 5(c)) and ion (Fig. 3(a)). It is remarked that introduction of dynamic ions in this case has led to double conduction band in the middle and has significantly increased the electron population in conduction band. Moreover, small wavelength excitations appear naturally as phonon assisted plasmon excitation in this system.

\section{Conclusion}

In this research we developed a new theory of plasmon band structure in plasmas. We used the linearized Schr\"{o}dinger-Poisson model and the pseudoforce method to study the energy band structure of plasmon excitations in different plasmas with arbitrary degree of degeneracy. Our study reveals the fundamental role played by dynamic ions on the band structure and available energy levels for electrons in the conduction bands. Many valence and conduction-like structure is found to be present in the band structure of the electron-ion plasma with fascinating resemblance to band structure of solids. We also investigated the effect of charge screening effect on the energy band and revealed its many important impact on quantum dense plasmas. The significant effects of ion charge state and electron chemical potential on the plasmon excitations was revealed and band structure of different pair plasmas were studied as well.

\section{Data Availability}

The data that support the findings of this study are available from the corresponding author upon reasonable request.

\end{document}